\documentclass[a4paper,11pt]{article}
\usepackage{pos}
\usepackage{wrapfig}
\usepackage{enumitem}
\usepackage{graphicx}
\usepackage{titlesec}
\usepackage{caption}
\usepackage{subcaption}
\usepackage{booktabs}
\usepackage{lineno}
\usepackage[capitalize]{cleveref}

\graphicspath{{figures/}}

\title{The Underground Muon Detector of AugerPrime: Status and Performance}
\ShortTitle{Status and {P}erformance of the {U}nderground {M}uon {D}etector}

\author*[a,b]{Joaquín de Jesús }
\onbehalf{for the Pierre Auger Collaboration$^c$}

\affiliation[a]{Instituto de Tecnología y Detección en Astropartículas (CNEA, CONICET, UNSAM),\\
  Buenos Aires, Argentina}

\affiliation[b]{Karlsruhe Institute of Technolgy (KIT), Institute for Astroparticle Physics, Karlsruhe, Germany}

\affiliation[c]{Observatorio Pierre Auger, Av.\ San Mart{\'\i}n Norte 304, 5613 Malarg\"ue, Argentina\\
Full author list: \normalfont{\rm\url{ https://www.auger.org/archive/authors_icrc_2025.html}}}

\emailAdd{spokespersons@auger.org}

\abstract{As part of the upgrade of the Pierre Auger Observatory, known as AugerPrime, the Underground Muon Detector is being installed in the low-energy extension of the Surface Detector, allowing for a direct measurement of the muonic component of air showers produced by ultra-high-energy cosmic rays with energies between $10^{16.8}$ and $10^{19}$\,eV.
The detector consists of an array of 30\,m$^2$ plastic scintillator detectors buried 2.3\,m underground near a water-Cherenkov detector.
Two modes of operation are implemented to achieve an extended dynamic range: the binary mode, conceived for low muon densities, and the calorimetric mode, designed for high muon densities.
In this contribution, we present the current status and improvements in the reconstruction and performance of this component of AugerPrime.}

\ConferenceLogo{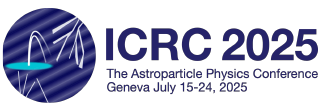}

\FullConference{%
The 39th International Cosmic Ray Conference (ICRC2025)\\
15 -- 24 July, 2025\\
Geneva, Switzerland}

\begin{document}
\maketitle

\section{Introduction}

Successfully operating for more than 20 years, the Pierre Auger Observatory is the largest facility dedicated to the detection of ultra-high-energy cosmic rays (UHECRs).
It employs a hybrid technique for the detection of extensive air showers (EAS) induced by UHECRs, combining a Surface Detector (SD) and a Fluoresence Detector (FD).
The SD, designed to measure the footprint of EAS at ground level, spans 3000 km$^2$ and is composed of an array of 1660 water-Cherenkov detectors (WCDs), spaced 1500 m apart (SD-1500) and arranged in a triangular grid.
The SD is complemented by the FD, which consists of 27 fluorescence telescopes arranged at five sites.
These telescopes allow for the measurement of the longitudinal development of EAS during nights with a low moon fraction and favourable weather conditions.

To extend the sensitivity of the Observatory to smaller energies, two nested, denser arrays are operative in a smaller part of the SD array.
The SD-750, with a spacing of 750 m between the WCDs, covers 23 km$^2$ and is fully efficient for energies above 10$^{17.3}$ eV.
The SD-433, with 433~m spacing, spans 1.9 km$^2$ and allows us to detect EAS above 10$^{16.8}$ eV.
To improve its sensitivity to the primary mass, the Observatory has recently been upgraded through the AugerPrime initiative~\cite{castellina2019augerprime}.
This enhancement includes the addition of scintillator and radio detectors atop the existing WCDs, the replacements of the WCD electronics and the installation of the Underground Muon Detector (UMD).
\section{The Underground Muon Detector}

The UMD, currently being installed within the SD-750 (UMD-750) and SD-433 (UMD-433) arrays, is designed to provide a direct measurement of the muonic component of EAS.
Each UMD station is positioned near a WCD and buried at a depth of 2.3 meters, shielding the detector from electromagnetic particles and establishing an effective energy threshold of approximately 1 GeV for vertical muons.
A UMD station consists of three detection modules, each covering an area of 10.5 m$^2$ of plastic scintillator. 
These modules are highly segmented, comprising 64 scintillator strips—each 400 cm long, 4 cm wide, and 1 cm thick—equipped with wavelength-shifting optical fibers. 
The light generated by muons passing through the strips is collected by the fibers and read out by an array of 64 silicon photomultipliers (SiPMs).
The UMD operates in slave mode, acquiring data only when triggered by the associated WCD.

To extend the dynamic range of the UMD, its modules operate in two complementary modes: the calorimetric (also known as ADC) mode and the binary mode.
The calorimetric mode, designed for high muon densities, treats the module as a
whole independent of detector segmentation. In this mode, the 64 SiPM signals are summed and subsequently amplified with high- and low-gain amplifiers. 
The amplified signals are digitized with
two analog-to-digital converters (ADCs) at a sampling time of 6.25 ns, producing two waveforms of 1024 samples.
The number of muons is then obtained by dividing the charge of these signals by the mean charge of a single vertical muon~\cite{uhecr_fal}.
On the other hand, the binary mode relies on detector segmentation and processes each of the 64 SiPM signals independently.
The output of each SiPM is processed by
a dedicated channel, producing a binary trace of 2048 samples, hence its name. 
In each sample, a ``1'' or ``0'' is recorded if the signal of the SiPM — after some electronic processing — is above or below a
discriminator threshold, respectively.
Muon signals in each bar are identified as a sequence of at least four consecutive ``1''s, which we refer to as a \emph{muon pattern}~\cite{botti2021calibration}.
The main observable for the reconstruction procedure of the muon lateral distribution function (LDF) in this mode is the activation number, $k$, which represents
the number of bars presenting a muon pattern in an air-shower event.

In this contribution, we present two improvements introduced in the  reconstruction procedure of the binary mode of the UMD-750. 
Their impact in the muon LDF fit is assessed using simulations.
In \cref{sec:core_umd}, we describe the new strategy to fit the shower core position using the data from the UMD.
In \cref{sec:cc}, we introduce a data-driven method to correct for the overcounting introduced by corner-clipping muons, which are inclined muons that generate muon patterns in two adjacent segments.

\section{Fitting the core position with the UMD}
\label{sec:core_umd}

The number of muons in a given detector in an air shower event follows a Poisson distribution with an expected value $\mu = \rho(r|\vec{p}) \, A \, \cos \theta$, where $\rho(r|\vec{p})$ is the muon density predicted by the muon LDF model, dependent on a set of parameters $\vec{p}$, and evaluated at the distance $r$ from the detector to the shower axis. 
Here, $A=10.5 \text{ m}^2$ denotes the detector area, and $\theta$ is the zenith angle of the shower. 

For a fixed $\mu$, the probability distribution of the observable $k$ is given by  (see Ref.~\cite{uhecr_fal} and references therein)
\begin{equation}
  \label{eq:prob_k}
  P(k|\mu) = {64 \choose k}  e^{-\mu(1+p_\text{cc})}(e^{\mu(1+p_\text{cc})/64} - 1)^k,
\end{equation}
where $p_\text{cc}$ is a quantity necessary to account for corner-clipping muons, as explained in \cref{sec:cc}.
For a measured $k$, \cref{eq:prob_k} represents the likelihood $L$ of the detector.
The event likelihood $\mathcal{L}$ is defined as the product of individual detector likelihoods, and its negative logarithm is minimized to obtain the best-fit LDF parameters $\vec{p}$.
Due to the sparse detector spacing, the LDF slope cannot be reliably fitted in most events and is therefore fixed to a parameterized mean value.
Only the muonic shower size $\rho(450)$—the LDF evaluated at 450~m and a proxy for the total number of muons—is fitted.
The reference distance of 450~m is chosen to minimize the fluctuations between the signal predicted by the average LDF used in the reconstruction and the actual, event-specific LDFs, which vary due to intrinsic shower-to-shower fluctuations.

The arrival direction and core position, needed to compute shower-plane distances for the UMD, are extracted from the SD reconstruction.
Until now, the UMD muon LDF has been fitted with the core fixed to the SD value, which has two main drawbacks.
First, the uncertainty in the core is not propagated to $\rho(450)$, leading to underestimated uncertainties.
Second, fixing the core can bias the fit when a detector lies close to the core, where the LDF rises steeply and small shifts in distance $r$ cause large changes in the expected signal.
To avoid this, the core position should be left free in the UMD fit.

To ensure a good core reconstruction, a fiducial quality cut is applied in the SD reconstruction, which requires that all neighboring stations of the station with the largest signal are functioning properly~\cite{sd_reco}.
Such condition could be translated to the UMD modules, allowing to leave the core free during the UMD fit.
However, this would come at the cost of losing too many events, since the UMD array was growing at varying paces through the years, with deployment still ongoing.
To overcome this, a different strategy was employed: we leave the core as a free parameter but we add an extra factor in the event likelihood that penalizes core positions that are too far from the SD core.
In this way, the fiducial cut can be avoided and the core fited in every event.
To this aim, we model the distribution of the UMD core $(x, y)$ as a bivariate Gaussian whose expected value corresponds to $(x_\text{SD}, y_\text{SD})$, the core position obtained by the SD, and whose covariance matrix is given by the covariance matrix of the core coordinates obtained in the SD fit.
When taking the logarithm of this term, a $\chi^2$-like term is obtained.
The event log-likelihood now reads
\begin{equation}
  \label{eq:global_like_core}
  -\ln \mathcal{L}(\vec{p}) = -\sum_i \ln L_i(\vec{p}) + \frac{1}{2}\chi^2,
\end{equation}
where $L$ corresponds to the likelihood of \cref{eq:prob_k} and the index $i$ goes over all the UMD modules in the event.
The $\chi^2$ is given by
\begin{equation}
\label{eq:chi2}
  \chi^2(x, y) = \frac{1}{1-\rho_\text{SD}^2} \left[ \left( \frac{x-x_\text{SD}}{\sigma_{x_\text{SD}}} \right)^2 + \left( \frac{y-y_\text{SD}}{\sigma_{y_\text{SD}}} \right)^2 - 2\frac{\rho_\text{SD}}{\sigma_{x_\text{SD}} \sigma_{y_\text{SD}}} (x-x_\text{SD}) (y-y_\text{SD}) \right],
\end{equation}
where $\sigma_{x_\text{SD}}$, $\sigma_{y_\text{SD}}$ are the errors in the SD core position and $\rho_\text{SD}$ corresponds to the correlation between the SD core coordinates.

\begin{figure}
    \centering
    \def\w{0.49}
    \includegraphics[width=\w\textwidth]{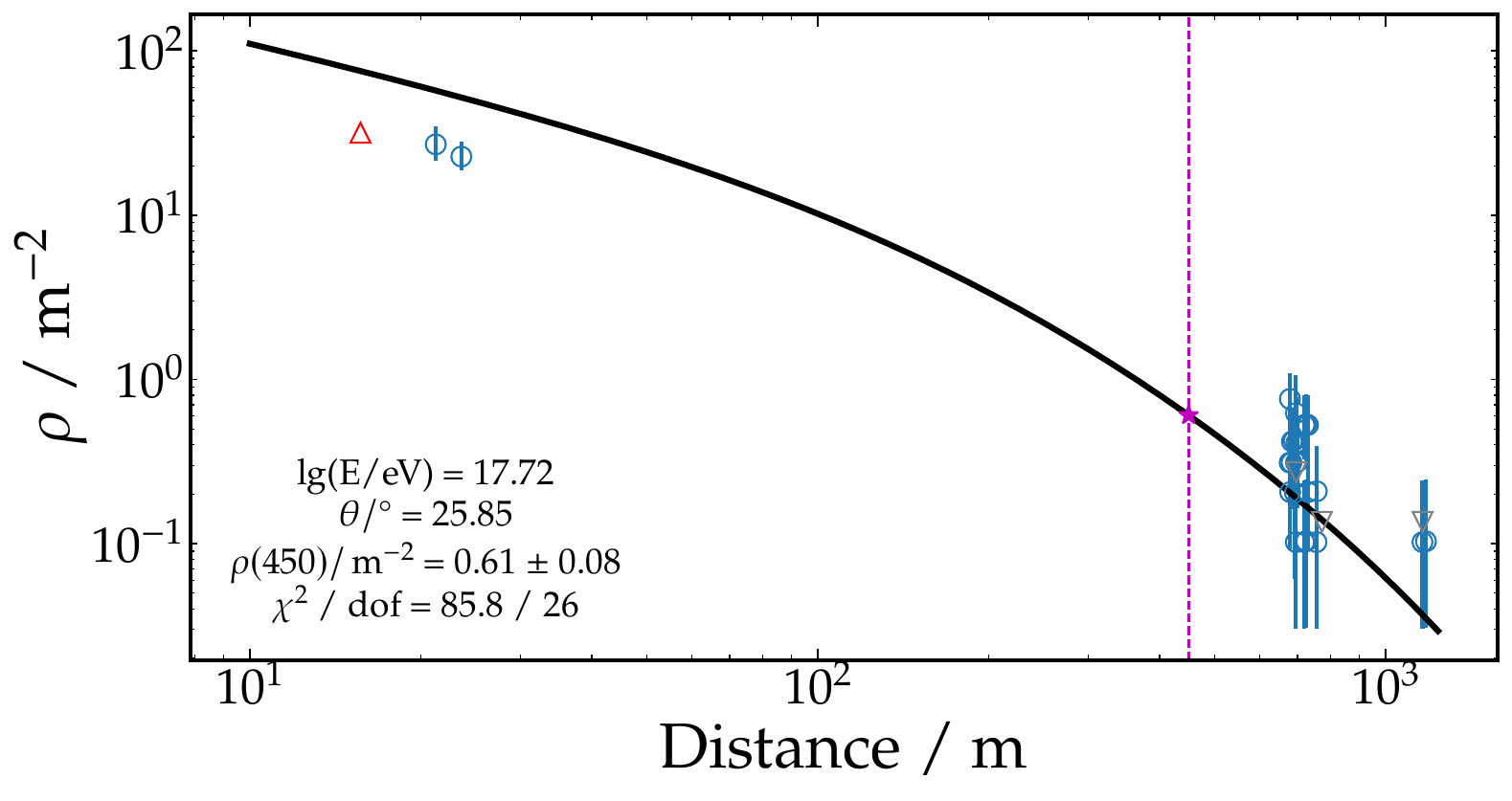}\hfill
    \includegraphics[width=\w\textwidth]{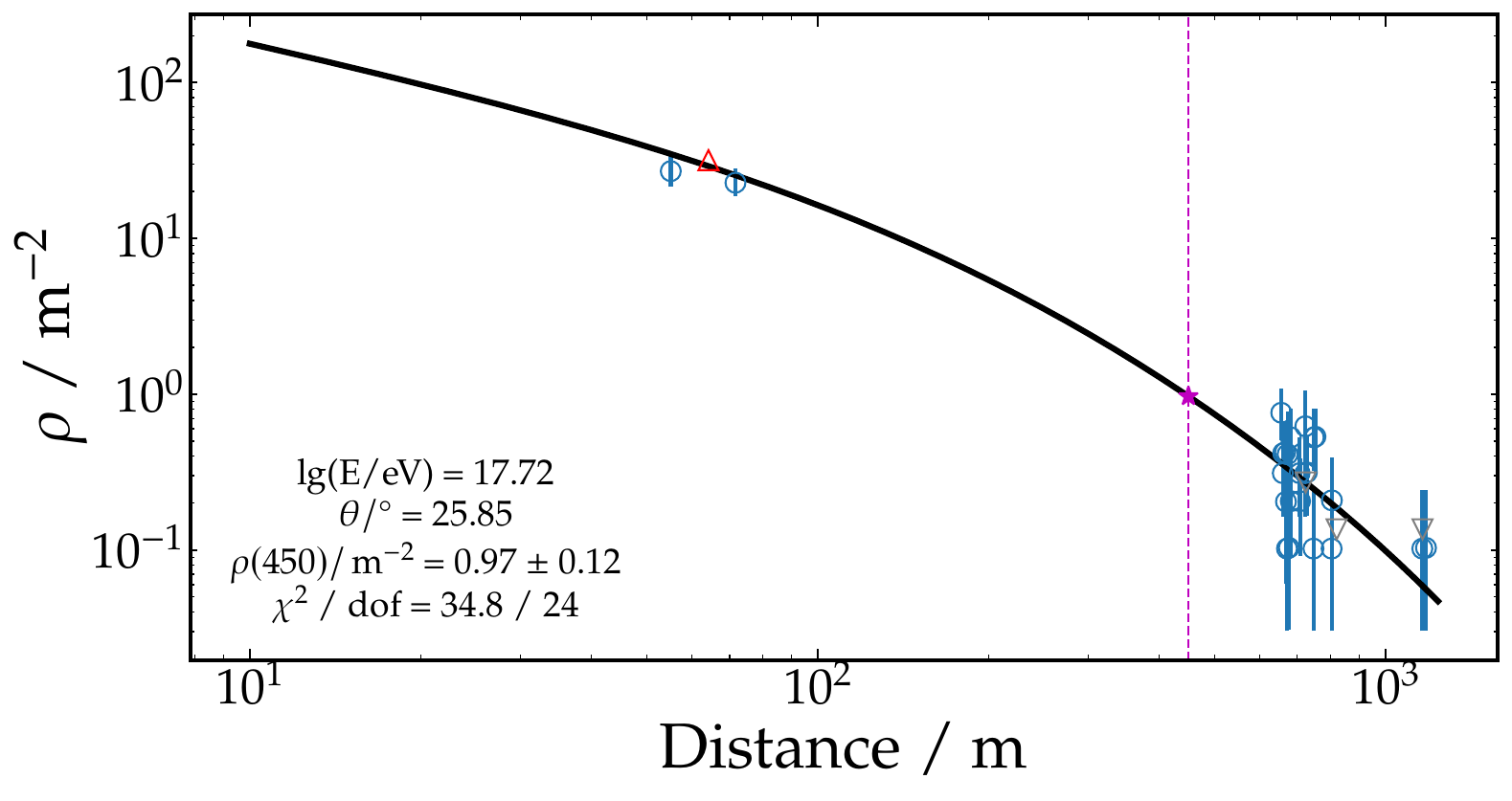}
    \caption{Muon LDF fit of a real event with the core fixed (left) and free (right). The dashed line marks the reference distance of 450 m. The red up-pointing triangle represents saturated detectors, whereas the grey down-pointing triangles indicate the upper limit of the muon density in detectors with zero muons. }
    \label{fig:fit_core_ldf}
\end{figure}

\begin{figure}
    \centering
    \def\w{0.55}
\includegraphics[width=\w\textwidth]{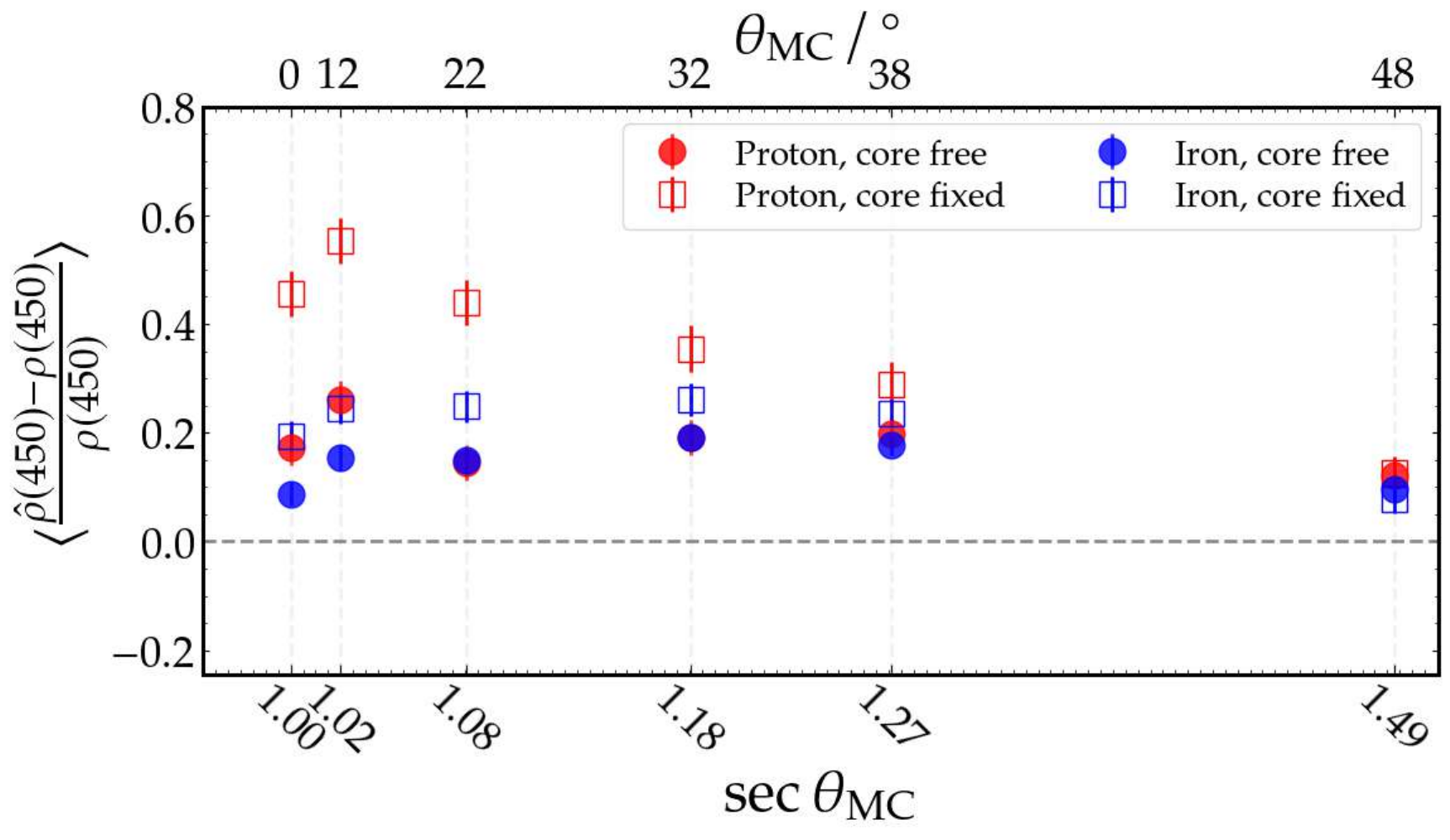}
    \caption{Bias in the muon shower size as a function of the secant of the Monte Carlo zenith angle. The case in which the core is left free (fixed) is indicated with full (empty) markers.}
    \label{fig:core_bias}
\end{figure}

In \cref{fig:fit_core_ldf}, we show the LDF fit of a real air shower with a core landing near a detector.
Fixing the core (left panel) results in a poor fit, with a reduced $\chi^2/\text{dof} = 3.3$ (p-value $= 2\times10^{-8}$) and an underestimated shower size of $\rho(450) = 0.61\,\text{m}^{-2}$.
Allowing the core to vary (right panel) yields a better fit, with $\chi^2/\text{dof} = 1.45$ (p-value = 7$\%$) and $\rho(450) = 0.97\,\text{m}^{-2}$.
A shift of just $\sim$48 meters between the SD and UMD cores leads to a significant change in shower size, highlighting the importance of treating the core position as a free parameter in the fit.
On average, the distance between the SD and UMD cores is 16 meters, and it is less than 47 meters for 95$\%$ of the events.
The impact of leaving the core free during the LDF fit on the shower size estimate was studied using simulations.
We used a library of CORSIKA~\cite{Heck_corsika} showers with protons and iron as primary species, EPOS-LHC~\cite{Pierog_2015} as the high-energy hadronic interaction model, an energy of $10^{17.5}$\,eV, and zenith angles of 0°, 12°, 22°, 32°, 38°, and 48°.
Detector simulations for both the WCDs and the UMD were performed using Offline~\cite{offline}, the official software framework of the Observatory.
The simulated signals were then passed through the full reconstruction pipeline, as is done with data.
To evaluate the scenario in which the core fit has the greatest impact, we forced all events to have a detector very close to the core by randomly placing each shower within a 100-meter-wide tile centered on a UMD station.
Each event was reconstructed with the core fixed to the SD core, and with the core free according to \cref{eq:global_like_core}.
The bias in each event was obtained by comparing the $\rho(450)$ retrieved by the LDF fit to the true value, obtained as the average of the true muon densities in a set of detectors placed at exactly 450 meters from the core in the simulations.
The mean bias with the core  fixed (empty markers) and free (solid markers) as a function of the zenith angle is shown in \cref{fig:core_bias} for both primaries.
When the core is fixed, the bias for proton (iron) can reach up to $\sim$50$\%$ ($\sim$20$\%$) for the most vertical zenith angles.
In contrast, the bias is lowered to $\sim 20\%$ ($\sim15\%$) when the core is free.
\section{Data-driven corner-clipping correction}
\label{sec:cc}
\begin{figure}
    \centering
    \def\w{0.36}
    \includegraphics[width=\w\textwidth]{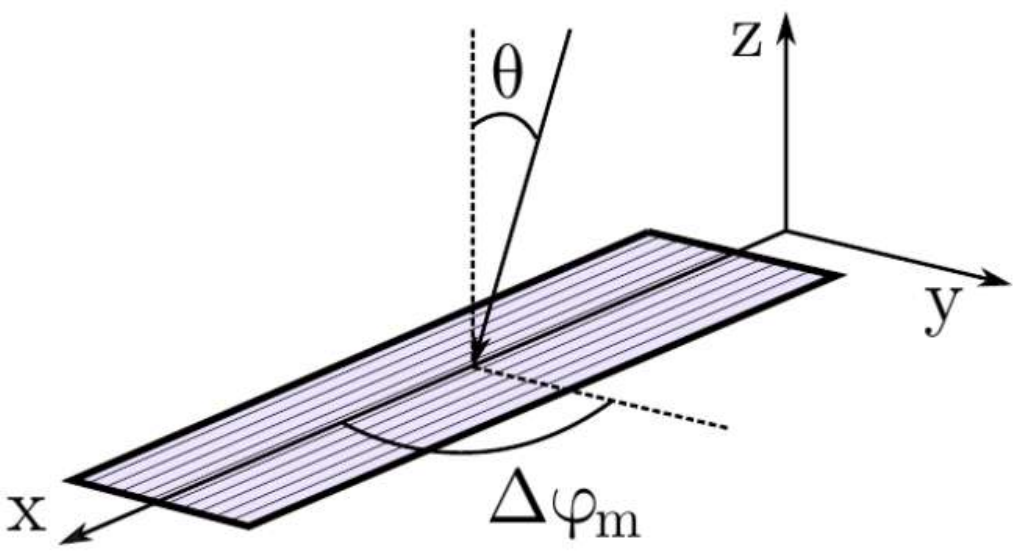}\hfill
    \includegraphics[width=\w\textwidth]{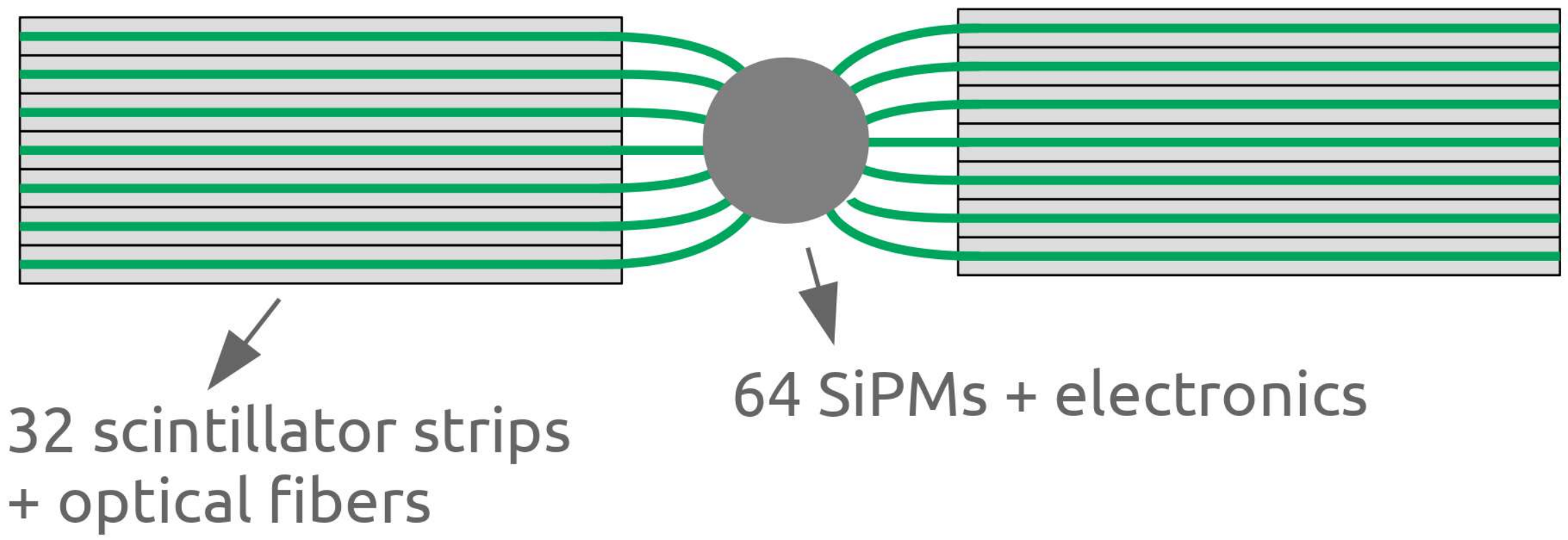}
    \caption{\emph{Left}: Geometry of the shower as seen by a UMD module. $\theta$ corresponds to the zenith angle and $\Delta \phi$ is the azimuth angle of the shower measured relative to the azimuth orientation of the detector. \emph{Right}: Sketch of a UMD module with its two halves.}
    \label{fig:cc_geom}
\end{figure}

Corner-clipping muons are inclined muons that can generate muon patterns in adjacent bars.
They are a source of overcounting and, if not properly accounted for, introduce a bias in the muon estimator that increases with the zenith angle $\theta$, and as $\Delta\phi$ — the azimuth angle as observed by the UMD module — becomes more perpendicular to the detector, see left panel of \cref{fig:cc_geom}.
To correct for this effect, the previous approach consisted of parameterizing the bias as a function of the zenith and azimuth angles of the shower, using air shower and full detector simulations.
This simulation-based correction was then applied to the data~\cite{pmtpaper}.
Recently, a new method was developed to account for the corner-clipping effect in a data-driven way~\cite{uhecr_cc}.
This method consists of estimating the probability that a single muon produces a signal in two adjacent bars — a quantity we refer to as the single-muon corner-clipping probability, $p_\text{cc}$.
Assuming that all muons arriving at a detector are independent, the number of corner-clipping muons in a detector can be modeled as a binomial process with a success probability of $p_\text{cc}$.
Under this assumption, the expected number of muons in a detector, $\mu$, is replaced by $\mu(1 + p_\text{cc})$ to account for the larger number of activated bars generated by corner-clipping muons, leading to \cref{eq:prob_k}~\cite{uhecr_cc}.

Considering that when a single muon is injected into one panel of the detector (see right panel of \cref{fig:cc_geom}), either one or two neighboring bars exhibit a muon pattern\footnote{We neglect inefficiencies—namely, cases where a muon fails to produce a muon pattern—based on the high efficiency of 98.5$\%$ measured with a single bar~\cite{botti2021calibration}.}, we can define $p_\text{cc}$ as
\begin{equation}
\label{eq:pcc}
p_\text{cc} = \frac{N_\text{cc}}{N_\text{cc} + N_{1}},
\end{equation}
where $N_\text{cc}$ and $N_1$ are the number of times two neighboring bars or one bar, respectively, were activated upon the injection of a single muon.
Since counting the number of events in which a detector had only one bar activated is straightforward, estimating $p_\text{cc}$ reduces to estimate $N_\text{cc}$.

\begin{figure}
    \centering
    \def\w{0.49}
    \includegraphics[width=\w\textwidth]{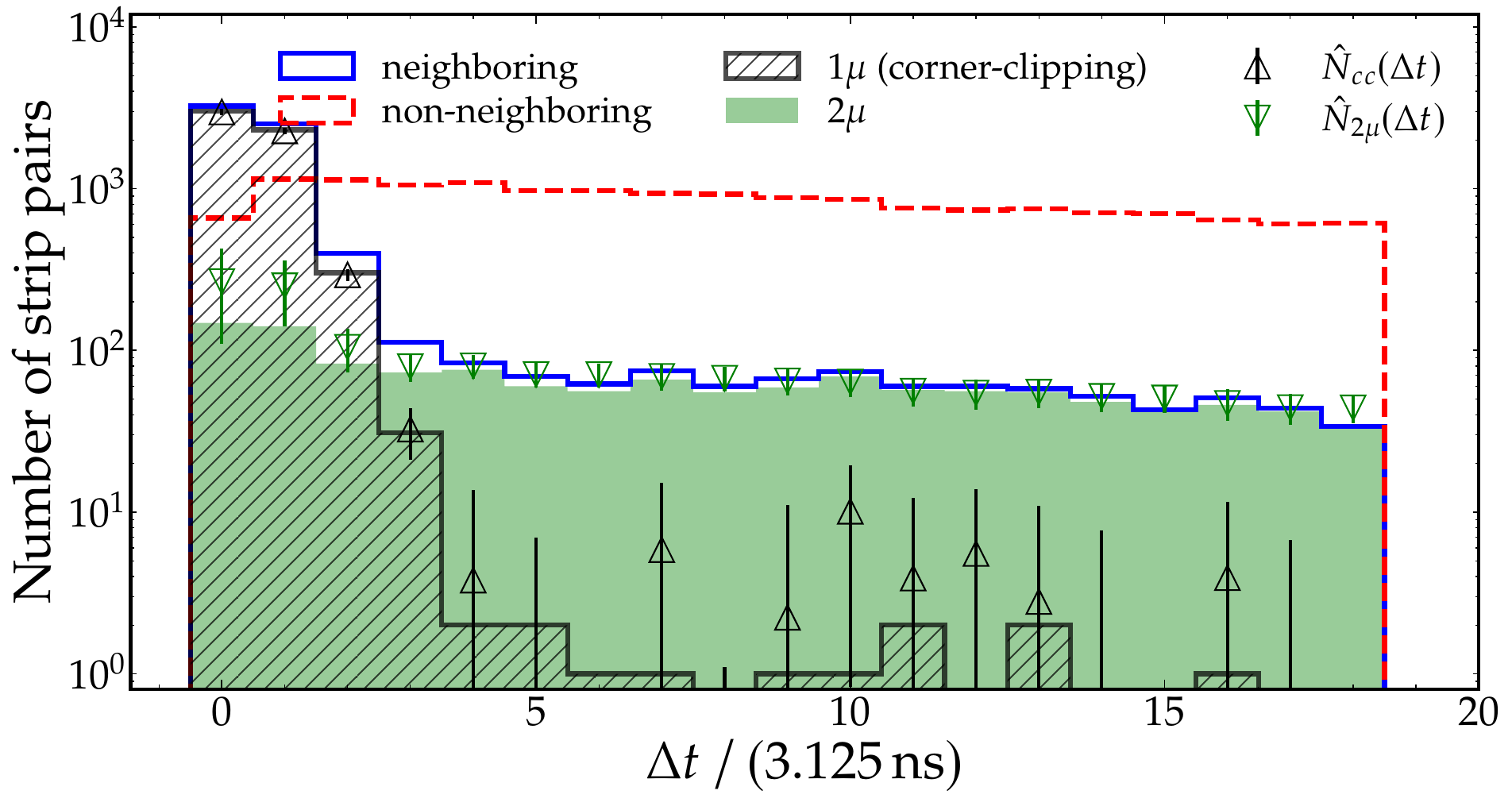}\hfill 
    \includegraphics[width=\w\textwidth]{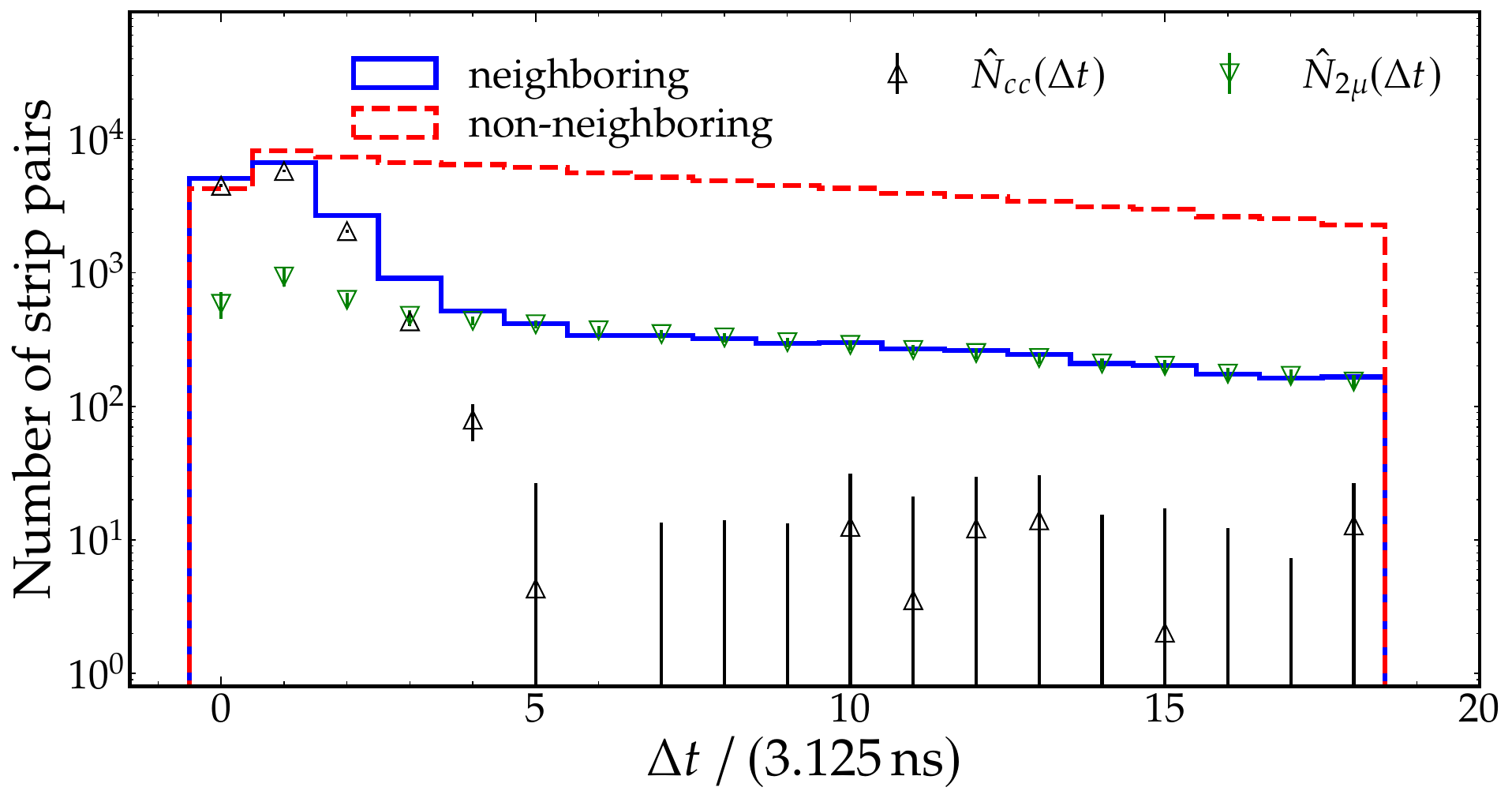}
    \caption{Distribution of $\Delta t$ for UMD-module panels with only two bars with a muon pattern for simulations (left) and data (right). The (non-)neighboring case is shown with a histogram with a solid (dashed) line. For simulations, the neighboring case is divided into the cases in which one (histogram with diagonal hatching) or two muons (green-filled histogram) were injected. The green down-pointing and black up-pointing triangular markers correspond to the estimators of the two-muon and single corner-clipping muon cases, respectively (see text for details).}
    \label{fig:dt_hist_sim_est}
\end{figure}

To estimate $N_\text{cc}$, we need to distinguish whether two neighboring bars are activated by a single corner-clipping muon or by two different muons.
For this reason, we use the timing information in detectors with only two activated bars.
In \cref{fig:dt_hist_sim_est}, we show the distribution of $\Delta t$, the absolute difference between the start times of the two muon patterns, for the neighboring (blue histogram with full lines) and non-neighboring (red histogram with dashed lines) cases, both for simulations (left panel) and data (right panel).
For simulations, a discrete library of CORSIKA showers with proton and iron primaries, energies of $10^{17.5}$, $10^{18}$, and $10^{18.5}$\,eV, and zenith angles of 0$^\circ$ , 12$^\circ$, 22$^\circ$, 32$^\circ$, 38$^\circ$, and 48$^\circ$ was used.
EPOS-LHC was employed as the high-energy hadronic interaction model.
In both data and simulations, a distinctive peak is observed in the neighboring case at small $\Delta t$, generated by corner-clipping muons producing two neighboring signals almost simultaneously.
In the simulations (left panel), cases where neighboring signals are produced by a corner-clipping muon or by two different muons are shown as histograms diagonally hatched in black and filled in green, respectively, confirming that the excess of neighboring signals at small $\Delta t$ is indeed due to corner-clipping muons.

In the absence of corner-clipping muons, the fraction of neighboring signals, $N_\text{neigh} / (N_\text{neigh} + N_\text{non-neigh})$, is expected to be 6.45$\%$, calculated as the probability of randomly selecting two neighboring bars out of 32.
Thus, for a given $\Delta t$, 6.45$\%$ of the neighboring signals is attributed to the two-muon case, while the remaining fraction is attributed to the corner-clipping case.
The estimated number of two-muon and corner-clipping muon events for each $\Delta t$ is shown as a green down-pointing triangle and a black up-pointing triangle, respectively.
In the simulations, the estimators closely follow the true Monte Carlo histograms, validating the method.
Summing over all the corner-clipping cases with $\Delta t / (3.125 \text{ ns}) < 6$, we obtain the total $N_\text{cc}$, which we subsequently insert in \cref{eq:pcc} to estimate $p_\text{cc}$.

\begin{figure}
    \centering
    \def\w{0.33}
\includegraphics[width=\w\textwidth]{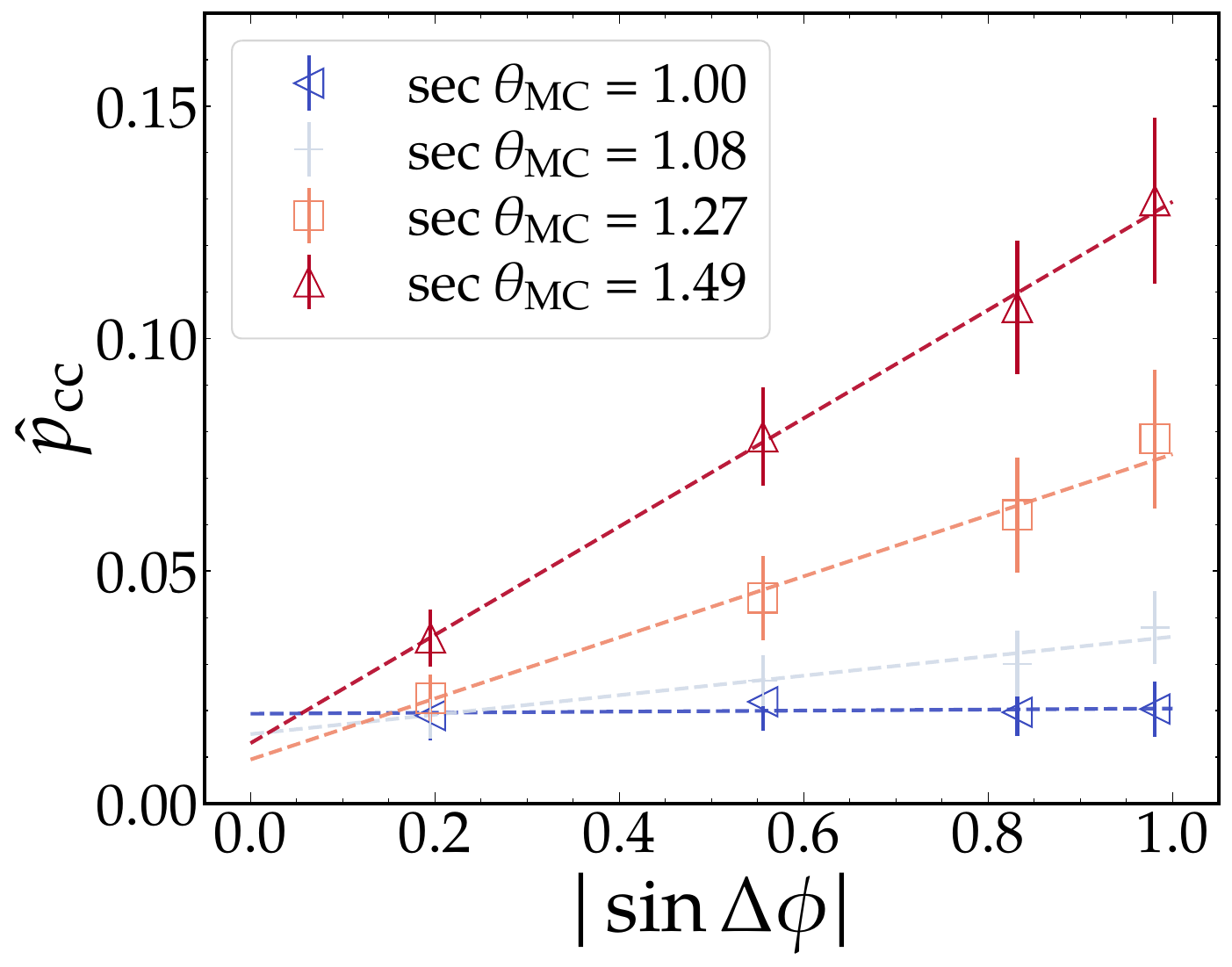}\hfill     \includegraphics[width=\w\textwidth]{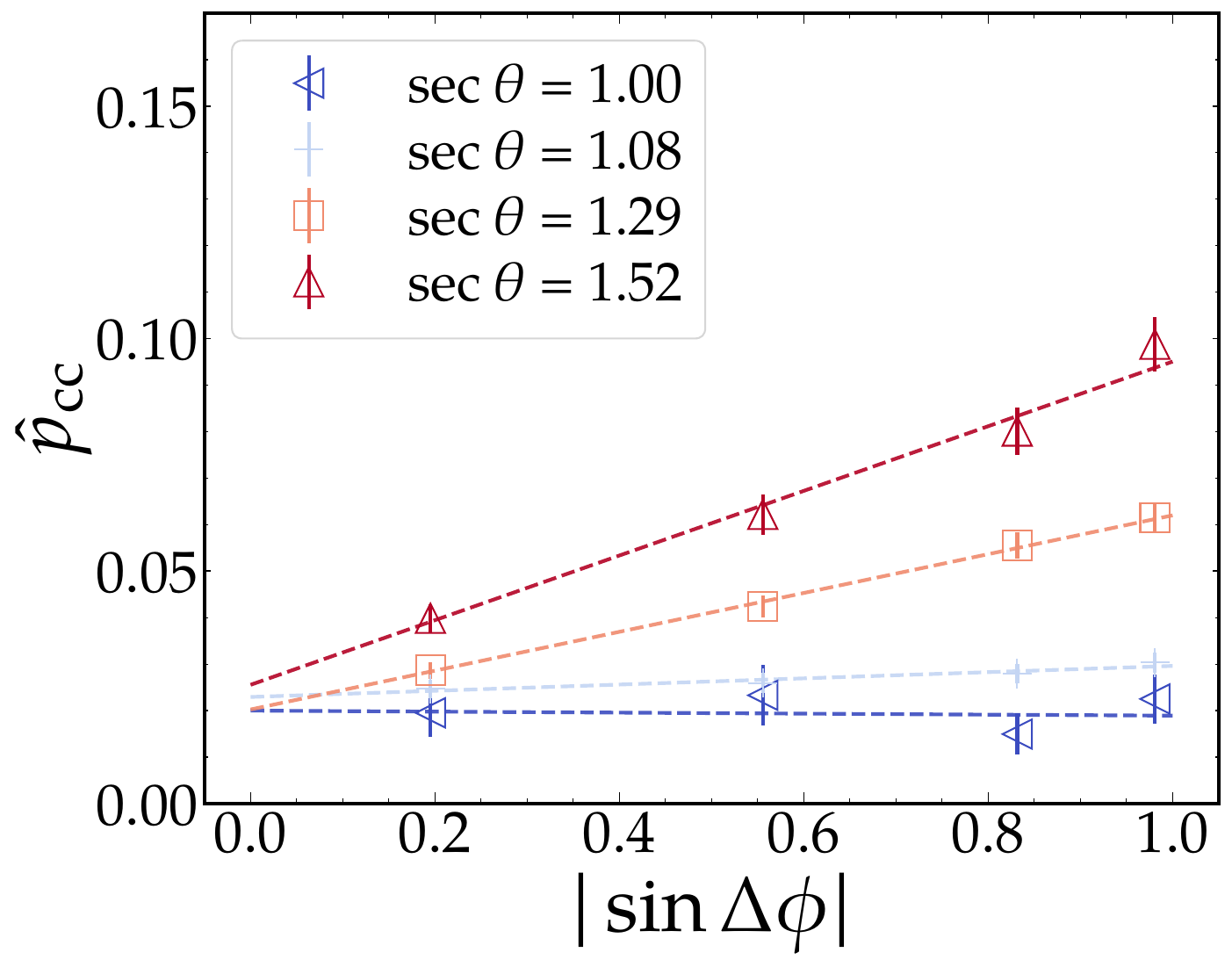}
\includegraphics[width=\w\textwidth]{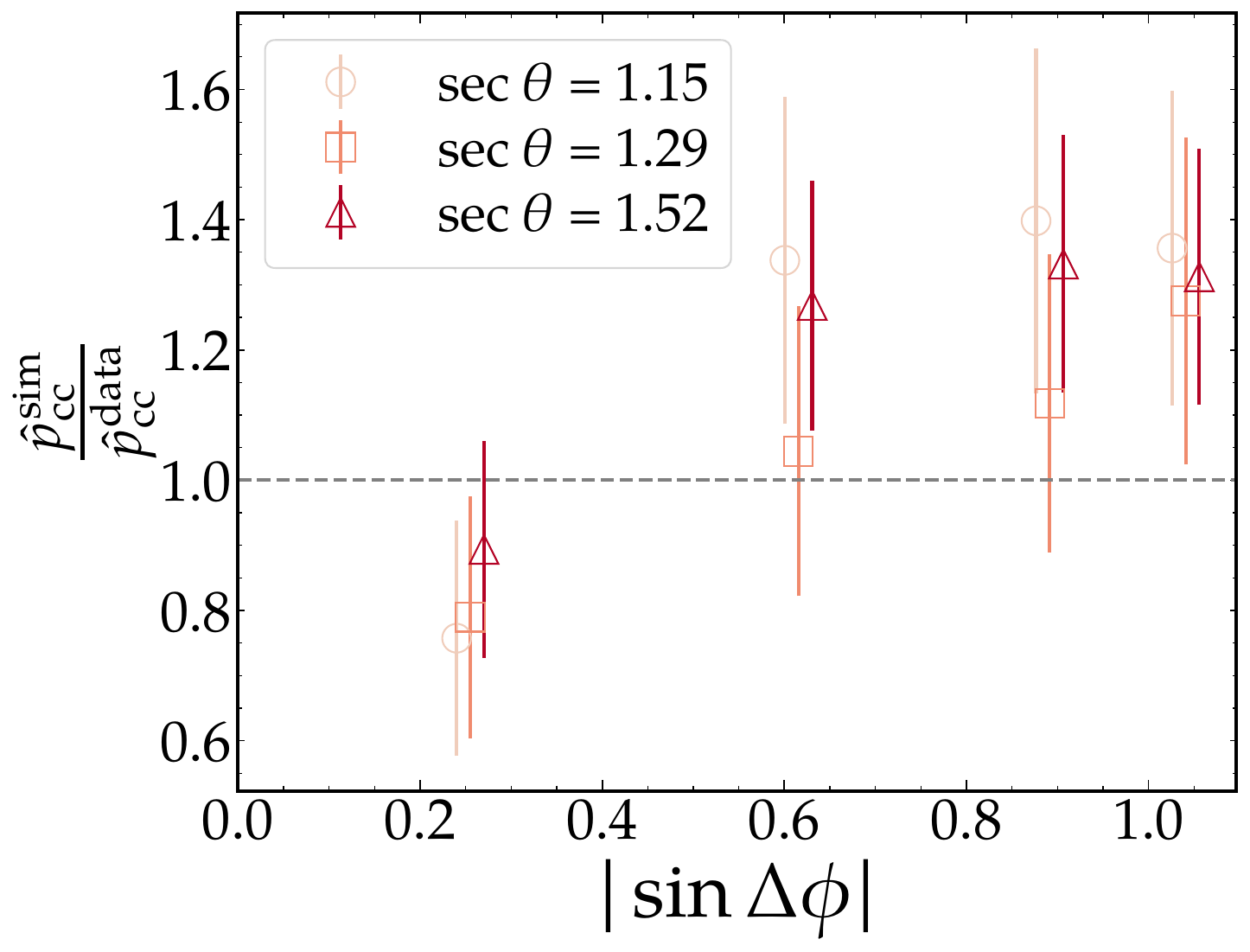}
    \caption{Estimated single-muon corner-clipping probability for simulations (left) and data (center). To ease visualization, only a subset of zenith angles are shown. The dashed lines indicate linear fits. For data, the secant of the center of the zenith angle bin is indicated in the legend. The right panel shows the ratio between the data and simulations for the most inclined zenith angles, where the discrepancy is largest. To improve clarity, the markers were slightly shifted horizontally to distinguish between the different zenith angles.}
    \label{fig:pcc_fits}
\end{figure}

By applying this method in bins of $\theta$ and $\Delta \phi$, the angular dependence of $p_\text{cc}$ can be obtained.
The results are shown in \cref{fig:pcc_fits} for simulations (left) and data (center).
The same qualitative behavior is observed in both cases, with the values for simulations being larger than those for data, particularly at the most inclined zenith angles (see right panel of \cref{fig:pcc_fits}), indicating that the corner-clipping effect is overestimated in the simulations.
As expected, $p_\text{cc}$ increases with $\theta$, and for a fixed $\theta$, it increases as $\Delta \phi$ approaches
90$^\circ$. 
For each $\theta$, we parameterize  the dependence of $p_\text{cc}$ by fitting a linear model in $|\sin\Delta\phi|$, displayed as dashed lines in the figure.
This result represents the first observation and quantification of the corner-clipping effect in data.

The difference between data and simulations likely stems from simplifications in the detector simulation.
For instance, the same scintillator-fiber-electronics response—tuned to limited laboratory data—is used for all simulated detectors, overlooking detector-to-detector variations in the field.
Consequently, different strips in the field might respond differently to identical energy depositions, leading to variations in efficiency and noise levels across the bars, all of which impact the estimated $p_\text{cc}$.

\begin{figure}
    \centering
    \def\w{0.55}
    \includegraphics[width=\w\textwidth]{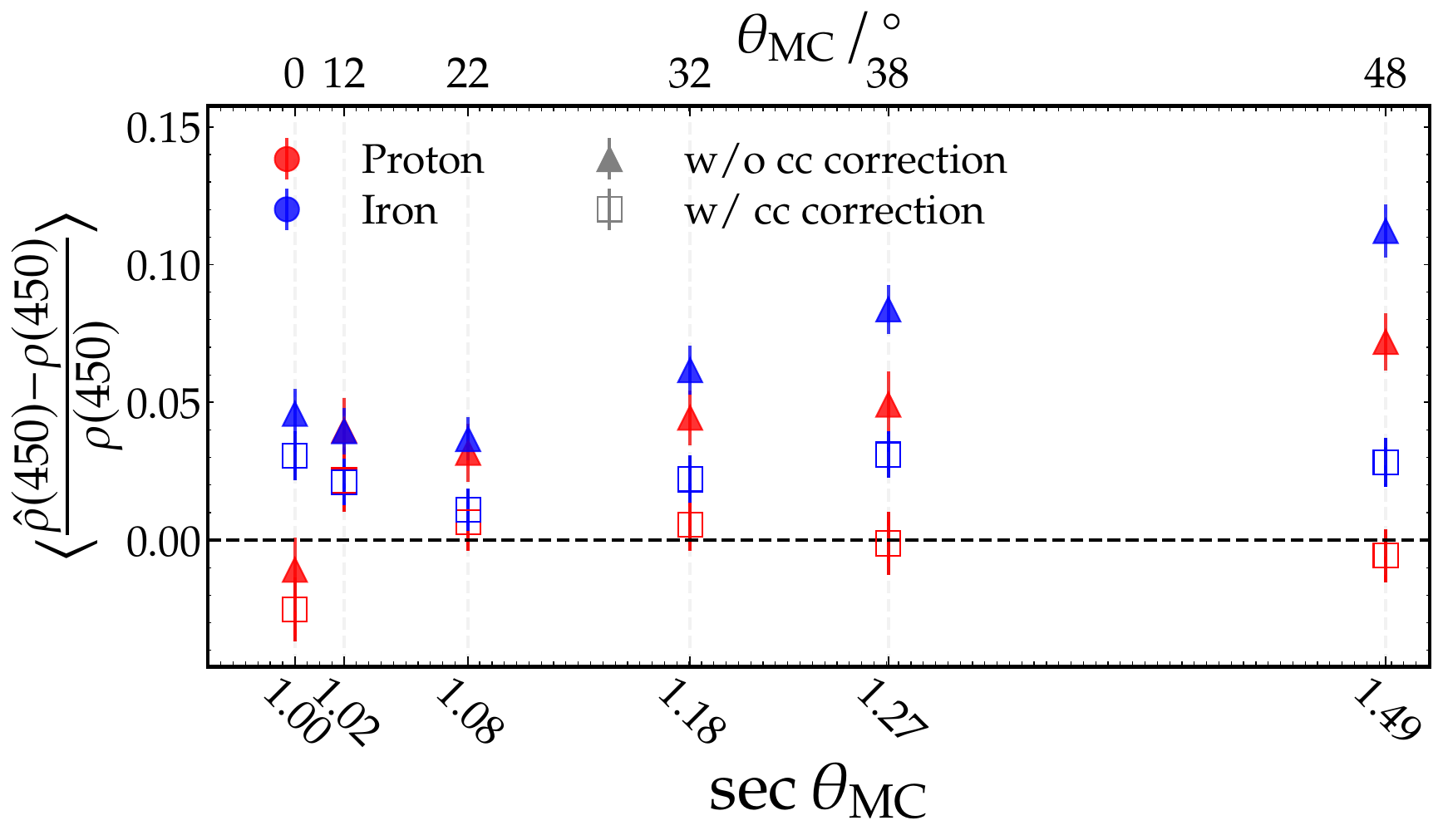}
    \caption{Bias in the muon shower size as a function of the secant of the Monte Carlo zenith angle with (empty squares) and without (full triangles) the corner-clipping correction.}
    \label{fig:cc_rho450_bias}
\end{figure}

Lastly, we used the subset of simulations with energy 10$^{18}$ eV to assess whether the estimated $p_\text{cc}$ is useful for correcting the bias introduced by the corner-clipping effect.
The muon LDF of each shower was reconstructed with (\cref{eq:prob_k}, using the linear parameterizations from \cref{fig:pcc_fits}) and without (\cref{eq:prob_k} with $p_\text{cc} = 0$) the corner-clipping correction.
The mean bias in the muonic shower size for the two cases, as a function of $\sec \theta_\text{MC}$, for proton (red) and iron (blue) is shown in \cref{fig:cc_rho450_bias}.
When no correction is applied, the bias increases with zenith angle as a consequence of corner-clipping muons.
In contrast, the bias remains flat, below $\sim 3\%$, when the correction is applied, indicating that the estimated $p_\text{cc}$ successfully reproduces the behavior of the bias.

\section{Summary}

In this contribution, we presented two improvements recently introduced in the reconstruction procedure of the binary mode of the Underground Muon Detector  of the Pierre Auger Observatory.

The muon shower size is estimated on an event-by-event basis by fitting a lateral distribution function (LDF) with fixed slope using a maximum likelihood approach.
We introduced a new strategy that adds a penalization term to the event likelihood, allowing the shower core position to be fitted directly within the UMD LDF reconstruction.
This removes the need for a fiducial cut, which would otherwise reduce statistics, and helps mitigate the bias in the shower size observed in events with a detector located very close to the core.

Additionally, we presented a data-driven method to correct for corner-clipping muons, based on the single-muon corner-clipping probability.
Simulations show that this probability successfully reproduces the behavior of the corner-clipping bias, enabling the correction of this effect—previously addressed only through simulations—using a purely data-driven approach.

\clearpage

\section*{The Pierre Auger Collaboration}
\small

\begin{wrapfigure}[8]{l}{0.11\linewidth}
\vspace{-5mm}
\includegraphics[width=0.98\linewidth]{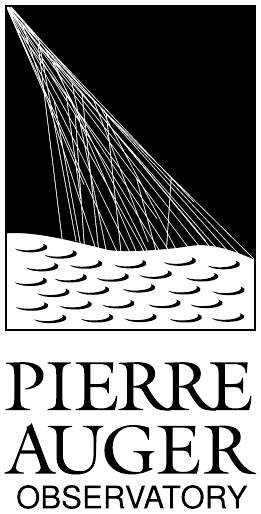}
\end{wrapfigure}
\begin{sloppypar}\noindent
% created on 2025-06-06
A.~Abdul Halim$^{13}$,
P.~Abreu$^{70}$,
M.~Aglietta$^{53,51}$,
I.~Allekotte$^{1}$,
K.~Almeida Cheminant$^{78,77}$,
A.~Almela$^{7,12}$,
R.~Aloisio$^{44,45}$,
J.~Alvarez-Mu\~niz$^{76}$,
A.~Ambrosone$^{44}$,
J.~Ammerman Yebra$^{76}$
% ...

\end{sloppypar}

\begin{center}
\rule{0.1\columnwidth}{0.5pt}
\raisebox{-0.4ex}{\scriptsize$\bullet$}
\rule{0.1\columnwidth}{0.5pt}
\end{center}

\vspace{-1ex}
\footnotesize
% created on 2025-06-06
% needs \usepackage{enumitem}
\begin{description}[labelsep=0.2em,align=right,labelwidth=0.7em,labelindent=0em,leftmargin=2em,noitemsep,before={\renewcommand\makelabel[1]{##1 }}]
\item[$^{1}$] Centro At\'omico Bariloche and Instituto Balseiro (CNEA-UNCuyo-CONICET), San Carlos de Bariloche, Argentina
\item[$^{2}$] Departamento de F\'\i{}sica and Departamento de Ciencias de la Atm\'osfera y los Oc\'eanos, FCEyN, Universidad de Buenos Aires and CONICET, Buenos Aires, Argentina
\item[$^{3}$] IFLP, Universidad Nacional de La Plata and CONICET, La Plata, Argentina
\item[$^{4}$] Instituto de Astronom\'\i{}a y F\'\i{}sica del Espacio (IAFE, CONICET-UBA), Buenos Aires, Argentina
\item[$^{5}$] Instituto de F\'\i{}sica de Rosario (IFIR) -- CONICET/U.N.R.\ and Facultad de Ciencias Bioqu\'\i{}micas y Farmac\'euticas U.N.R., Rosario, Argentina
\item[$^{6}$] Instituto de Tecnolog\'\i{}as en Detecci\'on y Astropart\'\i{}culas (CNEA, CONICET, UNSAM), and Universidad Tecnol\'ogica Nacional -- Facultad Regional Mendoza (CONICET/CNEA), Mendoza, Argentina
\item[$^{7}$] Instituto de Tecnolog\'\i{}as en Detecci\'on y Astropart\'\i{}culas (CNEA, CONICET, UNSAM), Buenos Aires, Argentina
\item[$^{8}$] International Center of Advanced Studies and Instituto de Ciencias F\'\i{}sicas, ECyT-UNSAM and CONICET, Campus Miguelete -- San Mart\'\i{}n, Buenos Aires, Argentina
\item[$^{9}$] Laboratorio Atm\'osfera -- Departamento de Investigaciones en L\'aseres y sus Aplicaciones -- UNIDEF (CITEDEF-CONICET), Argentina
\item[$^{10}$] Observatorio Pierre Auger, Malarg\"ue, Argentina
\item[$^{11}$] Observatorio Pierre Auger and Comisi\'on Nacional de Energ\'\i{}a At\'omica, Malarg\"ue, Argentina
\item[$^{12}$] Universidad Tecnol\'ogica Nacional -- Facultad Regional Buenos Aires, Buenos Aires, Argentina
\item[$^{13}$] University of Adelaide, Adelaide, S.A., Australia
\item[$^{14}$] Universit\'e Libre de Bruxelles (ULB), Brussels, Belgium
\item[$^{15}$] Vrije Universiteit Brussels, Brussels, Belgium
\item[$^{16}$] Centro Brasileiro de Pesquisas Fisicas, Rio de Janeiro, RJ, Brazil
\item[$^{17}$] Centro Federal de Educa\c{c}\~ao Tecnol\'ogica Celso Suckow da Fonseca, Petropolis, Brazil
\item[$^{18}$] Instituto Federal de Educa\c{c}\~ao, Ci\^encia e Tecnologia do Rio de Janeiro (IFRJ), Brazil
\item[$^{19}$] Universidade de S\~ao Paulo, Escola de Engenharia de Lorena, Lorena, SP, Brazil
\item[$^{20}$] Universidade de S\~ao Paulo, Instituto de F\'\i{}sica de S\~ao Carlos, S\~ao Carlos, SP, Brazil
\item[$^{21}$] Universidade de S\~ao Paulo, Instituto de F\'\i{}sica, S\~ao Paulo, SP, Brazil
\item[$^{22}$] Universidade Estadual de Campinas (UNICAMP), IFGW, Campinas, SP, Brazil
\item[$^{23}$] Universidade Estadual de Feira de Santana, Feira de Santana, Brazil
\item[$^{24}$] Universidade Federal de Campina Grande, Centro de Ciencias e Tecnologia, Campina Grande, Brazil
\item[$^{25}$] Universidade Federal do ABC, Santo Andr\'e, SP, Brazil
\item[$^{26}$] Universidade Federal do Paran\'a, Setor Palotina, Palotina, Brazil
\item[$^{27}$] Universidade Federal do Rio de Janeiro, Instituto de F\'\i{}sica, Rio de Janeiro, RJ, Brazil
\item[$^{28}$] Universidad de Medell\'\i{}n, Medell\'\i{}n, Colombia
\item[$^{29}$] Universidad Industrial de Santander, Bucaramanga, Colombia
\item[$^{30}$] Charles University, Faculty of Mathematics and Physics, Institute of Particle and Nuclear Physics, Prague, Czech Republic
\item[$^{31}$] Institute of Physics of the Czech Academy of Sciences, Prague, Czech Republic
\item[$^{32}$] Palacky University, Olomouc, Czech Republic
\item[$^{33}$] CNRS/IN2P3, IJCLab, Universit\'e Paris-Saclay, Orsay, France
\item[$^{34}$] Laboratoire de Physique Nucl\'eaire et de Hautes Energies (LPNHE), Sorbonne Universit\'e, Universit\'e de Paris, CNRS-IN2P3, Paris, France
\item[$^{35}$] Univ.\ Grenoble Alpes, CNRS, Grenoble Institute of Engineering Univ.\ Grenoble Alpes, LPSC-IN2P3, 38000 Grenoble, France
\item[$^{36}$] Universit\'e Paris-Saclay, CNRS/IN2P3, IJCLab, Orsay, France
\item[$^{37}$] Bergische Universit\"at Wuppertal, Department of Physics, Wuppertal, Germany
\item[$^{38}$] Karlsruhe Institute of Technology (KIT), Institute for Experimental Particle Physics, Karlsruhe, Germany
\item[$^{39}$] Karlsruhe Institute of Technology (KIT), Institut f\"ur Prozessdatenverarbeitung und Elektronik, Karlsruhe, Germany
\item[$^{40}$] Karlsruhe Institute of Technology (KIT), Institute for Astroparticle Physics, Karlsruhe, Germany
\item[$^{41}$] RWTH Aachen University, III.\ Physikalisches Institut A, Aachen, Germany
\item[$^{42}$] Universit\"at Hamburg, II.\ Institut f\"ur Theoretische Physik, Hamburg, Germany
\item[$^{43}$] Universit\"at Siegen, Department Physik -- Experimentelle Teilchenphysik, Siegen, Germany
\item[$^{44}$] Gran Sasso Science Institute, L'Aquila, Italy
\item[$^{45}$] INFN Laboratori Nazionali del Gran Sasso, Assergi (L'Aquila), Italy
\item[$^{46}$] INFN, Sezione di Catania, Catania, Italy
\item[$^{47}$] INFN, Sezione di Lecce, Lecce, Italy
\item[$^{48}$] INFN, Sezione di Milano, Milano, Italy
\item[$^{49}$] INFN, Sezione di Napoli, Napoli, Italy
\item[$^{50}$] INFN, Sezione di Roma ``Tor Vergata'', Roma, Italy
\item[$^{51}$] INFN, Sezione di Torino, Torino, Italy
\item[$^{52}$] Istituto di Astrofisica Spaziale e Fisica Cosmica di Palermo (INAF), Palermo, Italy
\item[$^{53}$] Osservatorio Astrofisico di Torino (INAF), Torino, Italy
\item[$^{54}$] Politecnico di Milano, Dipartimento di Scienze e Tecnologie Aerospaziali , Milano, Italy
\item[$^{55}$] Universit\`a del Salento, Dipartimento di Matematica e Fisica ``E.\ De Giorgi'', Lecce, Italy
\item[$^{56}$] Universit\`a dell'Aquila, Dipartimento di Scienze Fisiche e Chimiche, L'Aquila, Italy
\item[$^{57}$] Universit\`a di Catania, Dipartimento di Fisica e Astronomia ``Ettore Majorana``, Catania, Italy
\item[$^{58}$] Universit\`a di Milano, Dipartimento di Fisica, Milano, Italy
\item[$^{59}$] Universit\`a di Napoli ``Federico II'', Dipartimento di Fisica ``Ettore Pancini'', Napoli, Italy
\item[$^{60}$] Universit\`a di Palermo, Dipartimento di Fisica e Chimica ''E.\ Segr\`e'', Palermo, Italy
\item[$^{61}$] Universit\`a di Roma ``Tor Vergata'', Dipartimento di Fisica, Roma, Italy
\item[$^{62}$] Universit\`a Torino, Dipartimento di Fisica, Torino, Italy
\item[$^{63}$] Benem\'erita Universidad Aut\'onoma de Puebla, Puebla, M\'exico
\item[$^{64}$] Unidad Profesional Interdisciplinaria en Ingenier\'\i{}a y Tecnolog\'\i{}as Avanzadas del Instituto Polit\'ecnico Nacional (UPIITA-IPN), M\'exico, D.F., M\'exico
\item[$^{65}$] Universidad Aut\'onoma de Chiapas, Tuxtla Guti\'errez, Chiapas, M\'exico
\item[$^{66}$] Universidad Michoacana de San Nicol\'as de Hidalgo, Morelia, Michoac\'an, M\'exico
\item[$^{67}$] Universidad Nacional Aut\'onoma de M\'exico, M\'exico, D.F., M\'exico
\item[$^{68}$] Institute of Nuclear Physics PAN, Krakow, Poland
\item[$^{69}$] University of \L{}\'od\'z, Faculty of High-Energy Astrophysics,\L{}\'od\'z, Poland
\item[$^{70}$] Laborat\'orio de Instrumenta\c{c}\~ao e F\'\i{}sica Experimental de Part\'\i{}culas -- LIP and Instituto Superior T\'ecnico -- IST, Universidade de Lisboa -- UL, Lisboa, Portugal
\item[$^{71}$] ``Horia Hulubei'' National Institute for Physics and Nuclear Engineering, Bucharest-Magurele, Romania
\item[$^{72}$] Institute of Space Science, Bucharest-Magurele, Romania
\item[$^{73}$] Center for Astrophysics and Cosmology (CAC), University of Nova Gorica, Nova Gorica, Slovenia
\item[$^{74}$] Experimental Particle Physics Department, J.\ Stefan Institute, Ljubljana, Slovenia
\item[$^{75}$] Universidad de Granada and C.A.F.P.E., Granada, Spain
\item[$^{76}$] Instituto Galego de F\'\i{}sica de Altas Enerx\'\i{}as (IGFAE), Universidade de Santiago de Compostela, Santiago de Compostela, Spain
\item[$^{77}$] IMAPP, Radboud University Nijmegen, Nijmegen, The Netherlands
\item[$^{78}$] Nationaal Instituut voor Kernfysica en Hoge Energie Fysica (NIKHEF), Science Park, Amsterdam, The Netherlands
\item[$^{79}$] Stichting Astronomisch Onderzoek in Nederland (ASTRON), Dwingeloo, The Netherlands
\item[$^{80}$] Universiteit van Amsterdam, Faculty of Science, Amsterdam, The Netherlands
\item[$^{81}$] Case Western Reserve University, Cleveland, OH, USA
\item[$^{82}$] Colorado School of Mines, Golden, CO, USA
\item[$^{83}$] Department of Physics and Astronomy, Lehman College, City University of New York, Bronx, NY, USA
\item[$^{84}$] Michigan Technological University, Houghton, MI, USA
\item[$^{85}$] New York University, New York, NY, USA
\item[$^{86}$] University of Chicago, Enrico Fermi Institute, Chicago, IL, USA
\item[$^{87}$] University of Delaware, Department of Physics and Astronomy, Bartol Research Institute, Newark, DE, USA
\item[] -----
\item[$^{a}$] Max-Planck-Institut f\"ur Radioastronomie, Bonn, Germany
\item[$^{b}$] also at Kapteyn Institute, University of Groningen, Groningen, The Netherlands
\item[$^{c}$] School of Physics and Astronomy, University of Leeds, Leeds, United Kingdom
\item[$^{d}$] Fermi National Accelerator Laboratory, Fermilab, Batavia, IL, USA
\item[$^{e}$] Pennsylvania State University, University Park, PA, USA
\item[$^{f}$] Colorado State University, Fort Collins, CO, USA
\item[$^{g}$] Louisiana State University, Baton Rouge, LA, USA
\item[$^{h}$] now at Graduate School of Science, Osaka Metropolitan University, Osaka, Japan
\item[$^{i}$] Institut universitaire de France (IUF), France
\item[$^{j}$] now at Technische Universit\"at Dortmund and Ruhr-Universit\"at Bochum, Dortmund and Bochum, Germany
\end{description}

\vspace{-1ex}
\footnotesize
% created on 2025-06-06
\section*{Acknowledgments}

\begin{sloppypar}
The successful installation, commissioning, and operation of the Pierre
Auger Observatory would not have been possible without the strong
commitment and effort from the technical and administrative staff in
Malarg\"ue. We are very grateful to the following agencies and
organizations for financial support:
\end{sloppypar}

\begin{sloppypar}
Argentina -- Comisi\'on Nacional de Energ\'\i{}a At\'omica; Agencia Nacional de
Promoci\'on Cient\'\i{}fica y Tecnol\'ogica (ANPCyT); Consejo Nacional de
Investigaciones Cient\'\i{}ficas y T\'ecnicas (CONICET); Gobierno de la
Provincia de Mendoza; Municipalidad de Malarg\"ue; NDM Holdings and Valle
Las Le\~nas; in gratitude for their continuing cooperation over land
access; Australia -- the Australian Research Council; Belgium -- Fonds
de la Recherche Scientifique (FNRS); Research Foundation Flanders (FWO),
Marie Curie Action of the European Union Grant No.~101107047; Brazil --
Conselho Nacional de Desenvolvimento Cient\'\i{}fico e Tecnol\'ogico (CNPq);
Financiadora de Estudos e Projetos (FINEP); Funda\c{c}\~ao de Amparo \`a
Pesquisa do Estado de Rio de Janeiro (FAPERJ); S\~ao Paulo Research
Foundation (FAPESP) Grants No.~2019/10151-2, No.~2010/07359-6 and
No.~1999/05404-3; Minist\'erio da Ci\^encia, Tecnologia, Inova\c{c}\~oes e
Comunica\c{c}\~oes (MCTIC); Czech Republic -- GACR 24-13049S, CAS LQ100102401,
MEYS LM2023032, CZ.02.1.01/0.0/0.0/16{\textunderscore}013/0001402,
CZ.02.1.01/0.0/0.0/18{\textunderscore}046/0016010 and
CZ.02.1.01/0.0/0.0/17{\textunderscore}049/0008422 and CZ.02.01.01/00/22{\textunderscore}008/0004632;
France -- Centre de Calcul IN2P3/CNRS; Centre National de la Recherche
Scientifique (CNRS); Conseil R\'egional Ile-de-France; D\'epartement
Physique Nucl\'eaire et Corpusculaire (PNC-IN2P3/CNRS); D\'epartement
Sciences de l'Univers (SDU-INSU/CNRS); Institut Lagrange de Paris (ILP)
Grant No.~LABEX ANR-10-LABX-63 within the Investissements d'Avenir
Programme Grant No.~ANR-11-IDEX-0004-02; Germany -- Bundesministerium
f\"ur Bildung und Forschung (BMBF); Deutsche Forschungsgemeinschaft (DFG);
Finanzministerium Baden-W\"urttemberg; Helmholtz Alliance for
Astroparticle Physics (HAP); Helmholtz-Gemeinschaft Deutscher
Forschungszentren (HGF); Ministerium f\"ur Kultur und Wissenschaft des
Landes Nordrhein-Westfalen; Ministerium f\"ur Wissenschaft, Forschung und
Kunst des Landes Baden-W\"urttemberg; Italy -- Istituto Nazionale di
Fisica Nucleare (INFN); Istituto Nazionale di Astrofisica (INAF);
Ministero dell'Universit\`a e della Ricerca (MUR); CETEMPS Center of
Excellence; Ministero degli Affari Esteri (MAE), ICSC Centro Nazionale
di Ricerca in High Performance Computing, Big Data and Quantum
Computing, funded by European Union NextGenerationEU, reference code
CN{\textunderscore}00000013; M\'exico -- Consejo Nacional de Ciencia y Tecnolog\'\i{}a
(CONACYT) No.~167733; Universidad Nacional Aut\'onoma de M\'exico (UNAM);
PAPIIT DGAPA-UNAM; The Netherlands -- Ministry of Education, Culture and
Science; Netherlands Organisation for Scientific Research (NWO); Dutch
national e-infrastructure with the support of SURF Cooperative; Poland
-- Ministry of Education and Science, grants No.~DIR/WK/2018/11 and
2022/WK/12; National Science Centre, grants No.~2016/22/M/ST9/00198,
2016/23/B/ST9/01635, 2020/39/B/ST9/01398, and 2022/45/B/ST9/02163;
Portugal -- Portuguese national funds and FEDER funds within Programa
Operacional Factores de Competitividade through Funda\c{c}\~ao para a Ci\^encia
e a Tecnologia (COMPETE); Romania -- Ministry of Research, Innovation
and Digitization, CNCS-UEFISCDI, contract no.~30N/2023 under Romanian
National Core Program LAPLAS VII, grant no.~PN 23 21 01 02 and project
number PN-III-P1-1.1-TE-2021-0924/TE57/2022, within PNCDI III; Slovenia
-- Slovenian Research Agency, grants P1-0031, P1-0385, I0-0033, N1-0111;
Spain -- Ministerio de Ciencia e Innovaci\'on/Agencia Estatal de
Investigaci\'on (PID2019-105544GB-I00, PID2022-140510NB-I00 and
RYC2019-027017-I), Xunta de Galicia (CIGUS Network of Research Centers,
Consolidaci\'on 2021 GRC GI-2033, ED431C-2021/22 and ED431F-2022/15),
Junta de Andaluc\'\i{}a (SOMM17/6104/UGR and P18-FR-4314), and the European
Union (Marie Sklodowska-Curie 101065027 and ERDF); USA -- Department of
Energy, Contracts No.~DE-AC02-07CH11359, No.~DE-FR02-04ER41300,
No.~DE-FG02-99ER41107 and No.~DE-SC0011689; National Science Foundation,
Grant No.~0450696, and NSF-2013199; The Grainger Foundation; Marie
Curie-IRSES/EPLANET; European Particle Physics Latin American Network;
and UNESCO.
\end{sloppypar}

\end{document}